\documentstyle[11pt,aaspp4]{article}

\slugcomment{To appear in {\it P. A. S. P.}}

\lefthead{Schneider et al.}
\righthead{Spectroscopy of SDSS Objects}

\begin{document}

\begin{center}
\plotfiddle{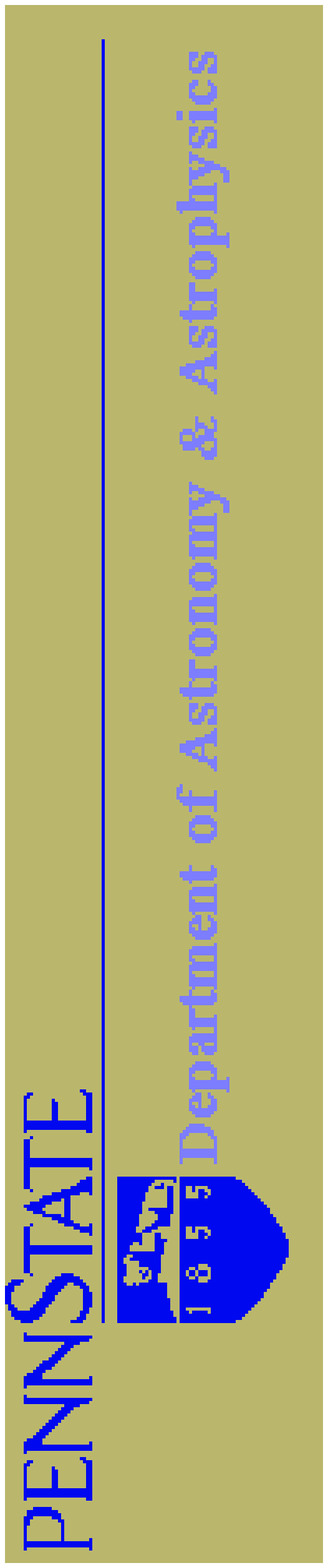}{0.5in}{-90}{50}{50}{-280}{200}
\end{center}

\rightskip 0pt \pretolerance=100

\title{\hbox{
The Low Resolution Spectrograph of the Hobby-Eberly Telescope II.}
Observations of Quasar Candidates from the Sloan Digital Sky Survey
\footnote{Based on observations obtained with the Sloan Digital
 Sky Survey, which is owned and operated by the Astrophysical Research
 Consortium.}$^,$\footnote{Based on observations obtained with the Hobby-Eberly
 Telescope, which is a joint project of the University of Texas at Austin,
 the Pennsylvania State University, Stanford University,
 Ludwig-Maximillians-Universit\"at M\"unchen, and Georg-August-Universit\"at
 G\"ottingen.}
}
\author{
D.P.~Schneider\altaffilmark{\ref{PennState}},
Gary~J.~Hill\altaffilmark{\ref{Texas}},
X.~Fan\altaffilmark{\ref{Princeton}},
L.W.~Ramsey\altaffilmark{\ref{PennState}},
P.J.~MacQueen\altaffilmark{\ref{Texas}},
D.W.~Weedman\altaffilmark{\ref{PennState}},
J.A.~Booth\altaffilmark{\ref{Texas}},
M.~Eracleous\altaffilmark{\ref{PennState}},
J.E.~Gunn\altaffilmark{\ref{Princeton}},
R.H.~Lupton\altaffilmark{\ref{Princeton}},
M.T.~Adams\altaffilmark{\ref{Texas}},
S.~Bastian\altaffilmark{\ref{FNAL}},
R.~Bender\altaffilmark{\ref{Munich}},
E.~Berman\altaffilmark{\ref{FNAL}},
J.~Brinkmann\altaffilmark{\ref{APO}},
I.~Csabai\altaffilmark{\ref{JHU}}$^,$\altaffilmark{\ref{Hungary}},
G.~Federwitz\altaffilmark{\ref{FNAL}},
V.~Gurbani\altaffilmark{\ref{FNAL}},
G.S.~Hennessy\altaffilmark{\ref{USNODC}},
G.M.~Hill\altaffilmark{\ref{Texas}},
R.B.~Hindsley\altaffilmark{\ref{USNODC}},
Z.~Ivezi\'c\altaffilmark{\ref{Princeton}},
G.R.~Knapp\altaffilmark{\ref{Princeton}},
D.Q.~Lamb\altaffilmark{\ref{Chicago}},
C.~Lindenmeyer\altaffilmark{\ref{FNAL}},
P.~Mantsch\altaffilmark{\ref{FNAL}},
C.~Nance\altaffilmark{\ref{Texas}},
T.~Nash\altaffilmark{\ref{FNAL}},
J.R.~Pier\altaffilmark{\ref{USNOAZ}},
R.~Rechenmacher\altaffilmark{\ref{FNAL}},
B.~Rhoads\altaffilmark{\ref{Texas}},
C.H.~Rivetta\altaffilmark{\ref{FNAL}},
E.L.~Robinson\altaffilmark{\ref{Texas}},
B.~Roman\altaffilmark{\ref{Texas}},
G.~Sergey\altaffilmark{\ref{FNAL}},
M.~Shetrone\altaffilmark{\ref{Texas}},
C.~Stoughton\altaffilmark{\ref{FNAL}},
M.A.~Strauss\altaffilmark{\ref{Princeton}},
G.P.~Szokoly\altaffilmark{\ref{FNAL}},
D.L.~Tucker\altaffilmark{\ref{FNAL}},
G.~Wesley\altaffilmark{\ref{Texas}},
J.~Willick\altaffilmark{\ref{Stanford}},
P.~Worthington\altaffilmark{\ref{Texas}},
and
D.G.~York\altaffilmark{\ref{Chicago}}
}

email addresses: dps@astro.psu.edu, hill@bento.as.utexas.edu,
fan@astro.princeton.edu

\newcounter{address}
\setcounter{address}{3}
\altaffiltext{\theaddress}{Department of Astronomy and Astrophysics, The
   Pennsylvania State University, University Park, PA 16802.
\label{PennState}}
\addtocounter{address}{1}
\altaffiltext{\theaddress}{Department of Astronomy, McDonald Observatory,
   University of Texas, Austin, TX~78712.
\label{Texas}}
\addtocounter{address}{1}
\altaffiltext{\theaddress}{Princeton University Observatory, Princeton,
   NJ 08544.
\label{Princeton}}
\addtocounter{address}{1}
\altaffiltext{\theaddress}{Fermi National Accelerator Laboratory, P.O. Box 500,
   Batavia, IL 60510.
\label{FNAL}}
\addtocounter{address}{1}
\altaffiltext{\theaddress}{Universit\"ats-Sternwarte,
   Scheinerstrasse~1, 81679~M\"unchen, Germany.
\label{Munich}}
\addtocounter{address}{1}
\altaffiltext{\theaddress}{Apache Point Observatory, P.O. Box 59,
   Sunspot, NM 88349-0059.
\label{APO}}
\addtocounter{address}{1}
\altaffiltext{\theaddress}{Department of Physics and Astronomy,
   Johns Hopkins University, 3701 University Drive, Baltimore, MD 21218.
\label{JHU}}
\addtocounter{address}{1}
\altaffiltext{\theaddress}{Department of Physics of Comples Systems,
   E\"otv\"os University, P\'azm\'ay P\'eter \hbox{s\'et\'any 1/A,}
   H-1117, Budapest, Hungary.
\label{Hungary}}
\addtocounter{address}{1}
\altaffiltext{\theaddress}{US Naval Observatory, 3450 Massachusetts Avenue NW,
   Washington, DC 20392-5420.
\label{USNODC}}
\addtocounter{address}{1}
\altaffiltext{\theaddress}{US Naval Observatory, Flagstaff Station,
   P.O. Box 1149, Flagstaff, AZ 86002-1149.
\label{USNOAZ}}
\addtocounter{address}{1}
\altaffiltext{\theaddress}{Department of Physics, Stanford University,
   Stanford, CA 94305.
\label{Stanford}}
\addtocounter{address}{1}
\altaffiltext{\theaddress}{Astronomy and Astrophysics Center, University of
   Chicago, 5640 South Ellis Avenue, Chicago, IL 60637.
\label{Chicago}}

\vbox{
\begin{abstract}
\rightskip 0pt \pretolerance=100\noindent
This paper describes spectra of quasar candidates
acquired during the commissioning phase of the
Low-Resolution Spectrograph of the Hobby-Eberly Telescope.  The
objects were identified as possible quasars from
multicolor image data from the Sloan Digital Sky Survey.  The ten sources
had typical $r'$ magnitudes \hbox{of 19--20},
except for one extremely red object
with~$r'$ of~$\approx$~23.  The data, obtained with exposure times between
10 and 25~minutes, reveal that the spectra of four candidates
are essentially featureless and are not quasars,
five are quasars with redshifts between~2.92 and~4.15 (including one Broad
Absorption Line quasar),
and the red source is a very late~M star or early~L dwarf.

\end{abstract}
}



%

\section{Introduction}

The Hobby-Eberly Telescope (HET), located at McDonald Observatory in
west Texas, is
the first optical/IR 8-m class telescope to employ a fixed altitude
(Arecibo-type)
design (\cite{rss94}; \cite{hill95}; \cite{lwr98}).
The spherical primary, consisting of 91 identical
hexagonal mirrors, is 11.1~m across and is
oriented 35$^{\circ}$ from the zenith; full azimuth
motion allows access to all declinations between~$-10^{\circ}$
and~$+72^{\circ}$.  During an observation the azimuth of the telescope
is fixed and objects are followed by a tracker assembly located
13.08~m above the primary, riding at the top of the telescope structure
(\cite{booth98}).  The tracker carries
a four-mirror corrector which delivers a 4$'$ diameter field of
view, and can follow an object
continuously for between 40~minutes and 2.5~hours, depending
on the source declination.
Only for a fraction of this time, however, does the 9.2-m diameter
entrance pupil fall entirely on the primary mirror; the minimum equivalent
aperture at the track extremes is 6.8-m, and the average equivalent
aperture for a ``typical" observation is approximately~8-m.

Groundbreaking for the HET occurred in March~1994, and the telescope
was dedicated on 8~October~1997.
The first HET facility instrument, the Marcario
Low Resolution Spectrograph (LRS; Hill~et~al.~1998a,b, 2000a; \cite{cobos98})
was installed
in the tracker in April~1999.  Commissioning of the LRS took place during
the dark time in April, May, and June~1999; this paper, along with
\cite{hill00b}, presents initial science results from these observations.

Spectra of ten high-redshift quasar candidates from the Sloan Digital Sky
Survey
(SDSS; \cite{gw95,SDSS96,york99}) were obtained with the LRS during the
Spring~1999
campaign.  The results demonstrated that although the image quality
of the HET primary had yet not reached design specifications, ten-minute
exposures of \hbox{$r' \approx$ 19--20} quasars were adequate to 
measure redshifts, and a twenty-five minute exposure of an
\hbox{$r' \approx 23$,} {$i' \approx 20.4$ L dwarf}
yielded an accurate stellar classification.

\section{Observations}

\subsection{Sloan Digital Sky Survey} \label{survtech}

The SDSS is using a CCD camera (\cite{gcam98}) on a
dedicated 2.5-m telescope (\cite{ws99}) at Apache Point Observatory,
New Mexico, to obtain images in five broad optical bands over
10,000~deg$^2$ of the high Galactic latitude sky centered approximately
on the North Galactic Pole.  The five filters (designated $u'$, $g'$,
$r'$, $i'$, and $z'$) cover the entire wavelength range of the CCD
response (\cite{fig96}).  Photometric calibration is provided by simultaneous
observations with a 20-inch telescope at the same site.  The
survey data processing software provides the astrometric and photometric
calibrations, as well as identification and characterization of objects
(\cite{jrp99,rhl99b}).

The high photometric accuracy of the SDSS images and the information
provided by the $z'$ filter (central wavelength of 9130~\AA ) makes the
SDSS data an excellent source for identification of high-redshift
quasar candidates.
If the redshift of a quasar exceeds~$\approx$~3.5, the combination of the
strong Ly~$\alpha$ line (typical observed equivalent width of 400~\AA )
and absorption produced by intervening neutral hydrogen (at $z$~=~4
approximately half of the radiation shortward of the Ly~$\alpha$ emission
line is absorbed) causes the optical colors of
high-redshift quasars to radically deviate (often by more than a
magnitude; see \cite{fan99,fan99a}) from the colors of stars.

Fan~et~al.~(1999) were able to identify~15 new quasars at redshifts larger
than~3.65 (including four with \hbox{$z > 4.5$}) from early SDSS commissioning
data; recent work (\cite{fan00a}) has increased the number of SDSS
high-redshift quasars to over~35.  During the course of these investigations,
observations of a number of quasar candidates have revealed a significant
number of objects cooler than M-type stars (\cite{fan00b}),
including the identification
of the first field methane dwarf (\cite{mas99}).

Quasar candidates were selected, using the multicolor technique
similar to that of Fan et al. (1999a),
from point sources in
two equatorial SDSS strips taken in
March~1999.  Both low (\hbox{$z < 3.5$}, from the ``$u^*g^*r^*$" diagrams)
and high
(\hbox{$z > 3.5$}, from the ``$g^*r^*i^*$" and ``$r^*i^*z^*$" diagrams) redshift
quasar candidates were chosen.
Since the photometric measurements of this commissioning data
have not yet been placed on the final SDSS system, we use the symbols
$u^*$, $g^*$, $r^*$, $i^*$, and~$z^*$ to indicate that the
photometry is similar but not identical to the final SDSS photometric system.

\subsection{Spectroscopy of Quasar Candidates} \label{survfields}

Spectra of ten of the SDSS quasar candidates were obtained with the LRS
between April and June~1999.  Details of the optical and mechanical
design and the performance of the LRS are provided in the companion paper
(\cite{hill00a});
below is a brief description of the
instrument as employed for the present observations.

The LRS is mounted in the Prime Focus Instrument Package, which
rides on the HET tracker.  The image scale \hbox{is 4.89$''$ mm$^{-1}$}
at the entrance aperture to the LRS; the observations of SDSS objects
were taken with long slits with widths of~2$''$ or~3$''$.  The dispersive
element was \hbox{a 300 line mm$^{-1}$} grism blazed \hbox{at 5500 \AA .}
The detector is a thinned, antireflection-coated
3072~$\times$~1024 Ford Aerospace CCD.  The pixel size is 15$\mu$m; the
scale on the detector is~0.25$''$~pixel$^{-1}$.  The CCD has a gain
\hbox{of 2.5 $e^-$ ADU$^{-1}$}
and a read noise of approximately~7~$e^-$.  The CCD was
\hbox{binned $ 2 \times 2$}
during readout of the SDSS observations; this produced a data frame size
\hbox{of 1568 $\times$ 512} and an image scale of~0.50$''$~pixel$^{-1}$.

LRS wavelength calibration was provided by Ne, Cd, and Ar comparison lamps.  The
wavelength calibration between 4400--10,700~\AA\ is well fit
(rms residuals of about a tenth of a pixel) by a fourth-order polynomial.
The dispersion ranges \hbox{from 4.00 \AA\ pixel$^{-1}$} at the blue end
\hbox{to 4.76 \AA\ pixel$^{-1}$} in the near infrared.  The resolution
with the~2$''$ slit is~20~\AA\ ($R$~=~300 at 6000~\AA ).

The observing conditions varied from nearly photometric to scattered cloud
cover.  Typically 86 of the 91 segments of the primary were in operation
during the data acquisition, and the image quality ranged from~1.6$''$~(FWHM)
to over~5$''$.  The exposure time per object ranged from~10
to~25~minutes.  The relative flux calibration was provided by observations
of spectrophotometric standards, usually those of \cite{og83}.  Absolute
spectrophotometric calibration was carried out
by scaling each spectrum so that
$i^*$ magnitudes synthesized from the spectra matched the SDSS photometric
measurements.

Six of the ten SDSS quasar candidates had interesting spectra; finding
charts are given in Figure~1.  The official names
for the sources are \hbox{SDSSp Jhhmmss.ss+ddmmss.s}, where the coordinate
equinox is~J2000.  For brevity, the objects will be referred to as
simply \hbox{SDSS hhmm+dd} throughout most of this paper.  The spectra of the
remaining four objects were basically featureless; given the signal-to-noise
ratio of these four spectra, one can state that they are not quasars or
late-type stars.

The spectra from 4500--9200~\AA\ for the six sources are displayed in
Figure~2 (The data have been binned so that there are approximately
two pixels per resolution element).
The data for all objects were taken through
the~2$''$~slit except for the spectrum of SDSS~1624$-$00 (3$''$~slit); the
spectrum of SDSS~1405$-$00 was acquired with an~OG515 blocking filter.

Prominent spectral features are labelled in the figure.
Four of the sources are definitely quasars with redshifts between~2.92
and~4.15 (also see \cite{fan00a} for observations of SDSS~1310-00
and SDSS~1447$-$00),
one
(SDSS~1347+00) is probably a quasar \hbox{at $z \approx 3.8$}, and one
(SDSS~1430+00) is a very cool star or substellar object.

\begin{figure}
\plotfiddle{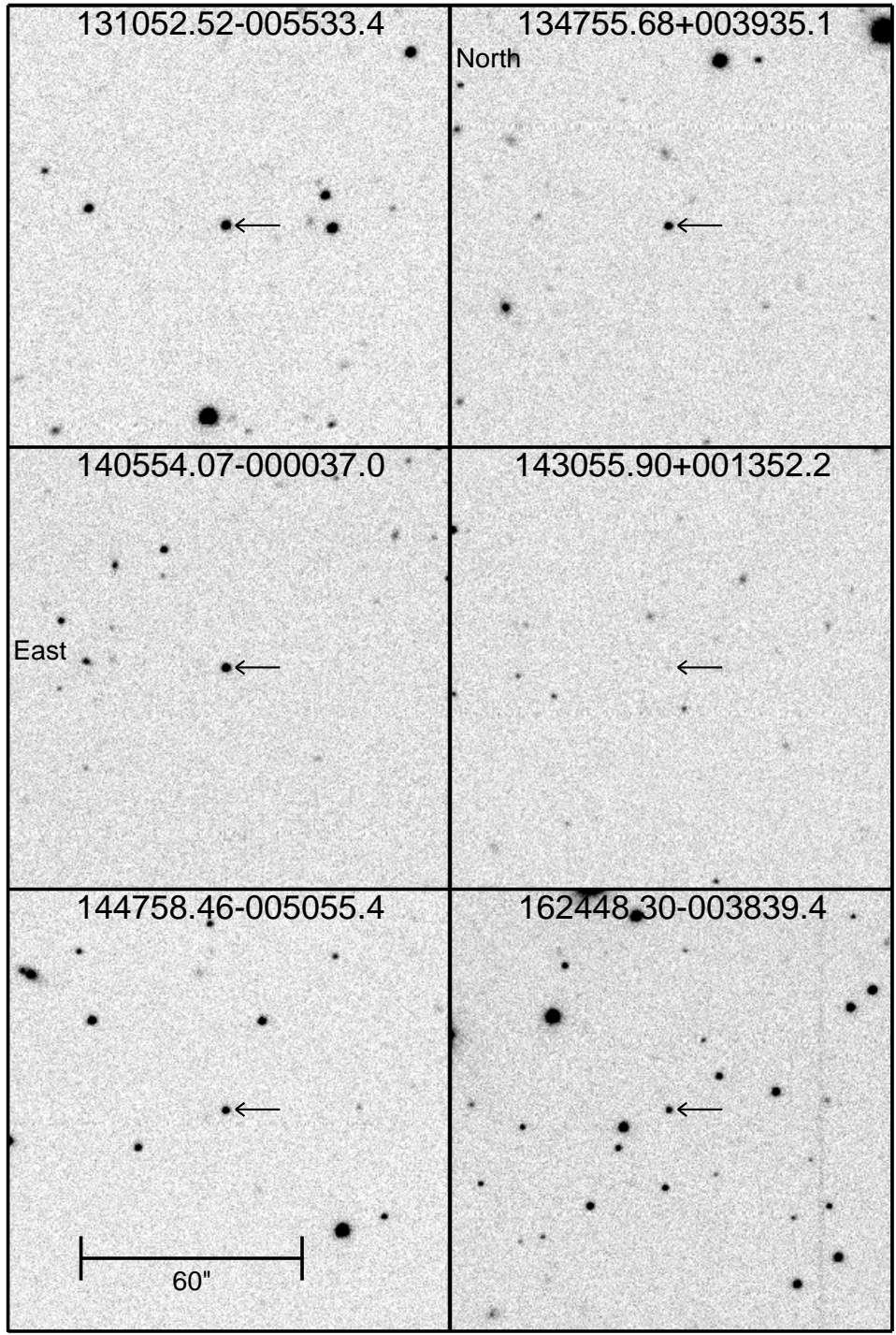}{7.5in}{0}{75}{75}{-170}{-30}
\caption{Finding charts for the six objects; the data were taken with
the SDSS imaging camera in the $r'$ filter.  The orientation of the charts
is north at the top and east to the left.  Each chart is 120$''$ on a side.}
\end{figure}

\begin{figure}
\plotfiddle{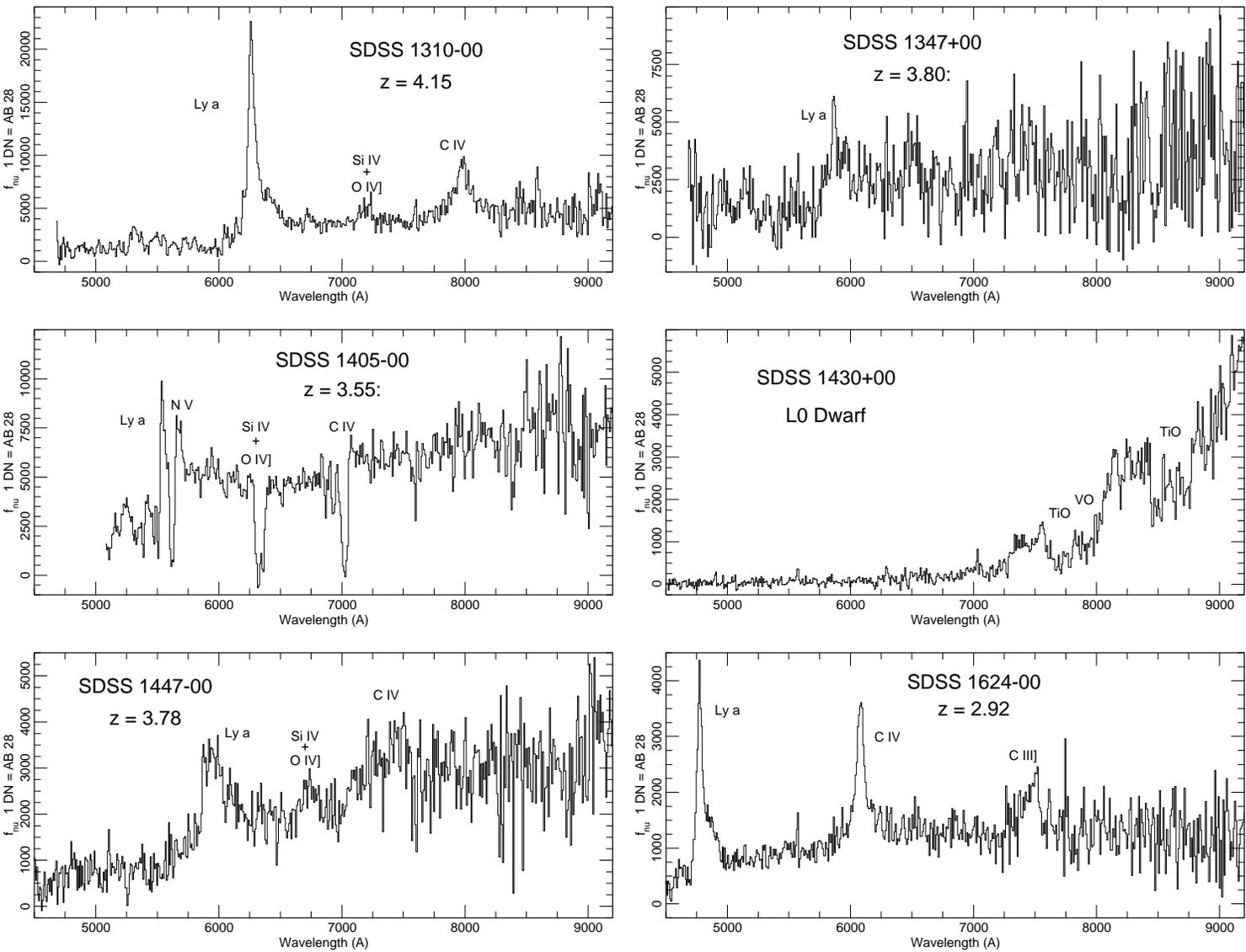}{7.5in}{180}{77}{77}{250}{600}
\caption{Spectra of the SDSS objects taken with the Low-Resolution
Spectrograph on the Hobby-Eberly Telescope.  Exposure times ranged
from 600~s to 1465~s.  The spectral resolution is 20~\AA\ except for
1624$-$00, where the resolution \hbox{is 30 \AA.}  The unit of flux is
\hbox{AB = 28.0} or \hbox{$5.75 \times 10^{-30}$ erg cm$^{-2}$
s$^{-1}$ Hz$^{-1}$.}}
\end{figure}

\section{Discussion}

A summary of the observations of the six SDSS sources are given in Table~1.
The table contains the object name, photometry with 1-$\sigma$ errors
from the SDSS data,
UT date and exposure time of LRS observation, and the redshift of the source.

The photometry in Table~1 is quoted in asinh magnitudes (\cite{lgs99a})
that are based on the $AB$ system of \cite{og83}.  Asinh magnitudes
are essentially identical to the standard definition of magnitudes
when the flux levels are well above zero; at low signal-to-noise
ratio asinh magnitudes are linear in flux and do not diverge
at zero (or negative) flux.  The zero flux levels
for the $u^*$, $g^*$, $r^*$, $i^*$, $z^*$ bands were set to
24.24, 24.91, 24.53, 23.89, and~22.47, respectively, for the SDSS
data reported here.  For example, the $g^*$ magnitude for
1420+00 (24.94) indicates that a small negative flux was measured
in this band.

Redshift determinations for three of the quasars (SDSS~1310$-$00,
SDSS~1447$-$00, and 
\hfil\break 
SDSS~1624$-$00) were quite straightforward;
emission lines other than the absorption-affected Ly~$\alpha$ line
could be used for this measurement.  The quasar SDSS~1405$-$00 is a
strong Broad Absorption Line (BAL) with absorption from Ly~$\alpha$,
N~V, \hbox{Si~IV+O~IV]}, and C~IV.  The BAL features have widths of
approximately~3000~km~s$^{-1}$ and have a redshift \hbox{of $\approx$
3.53;} we have assigned an approximate redshift of~3.55 to the quasar.
The spectrum of SDSS~1347+00, with its prominent emission line at
5900~\AA\ and continuum drop across the line, is very suggestive of a
redshift~$\approx$~3.8 object, but clearly a higher signal-to-noise
ratio spectrum of the source is required for a definitive answer.

Some basic properties of the five quasars are given in Table~2:
object name, redshift, color excess along line-of-sight
(from \cite{sfd98}), the Galactic
extinction corrected $AB$ magnitude at~1450~\AA\ in the rest frame of the
quasar, and the absolute $B$ magnitude (assuming
\hbox{$H_0$ = 50 km s$^{-1}$ Mpc$^{-1}$,} \hbox{$q_0$ = 0.5,}
and the continuum between~1450~\AA\ and~4400~\AA\ in the quasar rest frame
is a power law with an index of~$-0.5$).  In this cosmology 3C~273 has
\hbox{$M_B = -27.0$}.

The remaining object, SDSS~1430+00, is clearly a very late-type dwarf;
based on the classification scheme of \cite{kirk99}, we classify the
object as either late~M or early~L, with a best estimate of~L0
(also see \cite{fan00b}).

The results presented in this paper demonstrate that the HET/LRS can acquire,
track, and obtain spectra of~$\approx$~20${\rm th}$ magnitude objects.
The commissioning tests demonstrate that exposure times of the order of
ten minutes produce data of sufficient signal-to-noise ratio to determine
quasar redshifts and classify substellar objects at this brightness level.
The LRS will begin normal operations in Fall~1999; the observations presented
here are representative of the type of survey work that the LRS plans
to undertake in the future.

\begin{figure}
$$\vbox{\hsize=5truein
\halign{
# \hfil & 
\hfil # &
\hfil # &
\hfil # &
\hfil # &
\hfil # &
\hfil # \hfil &
\hfil # &
\hfil # \hfil \cr
\multispan9{\hfil TABLE 1. SDSS Objects Observed with the HET$^a$ \hfil}\cr
\noalign{\bigskip\hrule\smallskip\hrule\medskip}
&&&&&& \hfil Date \hfil & \hfil Exp \hfil \cr
\hfil Object$^b$ \hfil&\hfil $u^*$ \hfil&\hfil $g^*$ \hfil &
\hfil $r^*$ \hfil & \hfil $i^*$ \hfil & \hfil $z^*$ \hfil &
\hfil (1999) \hfil & \hfil (s) \hfil & $z$ \cr
\noalign{\medskip\hrule\bigskip}
SDSSp J131052.52$-$005533.4 & 23.54 & 20.87 & 18.84 & 18.79 & 18.86 & 12 Jun &
600 & 4.15 \cr
& $\pm$0.51 & $\pm$0.03 & $\pm$0.01 & $\pm$0.02 & $\pm$0.05 \cr
SDSSp J134755.68+003935.1   & 24.11 & 20.54 & 19.42 & 19.31 & 19.35 & 12 Jun &
600 & 3.80 \cr
& $\pm$0.19 & $\pm$0.02 & $\pm$0.01 & $\pm$0.02 & $\pm$0.05 \cr
SDSSp J140554.07$-$000037.0 & 24.10 & 20.16 & 18.81 & 18.54 & 18.34 & 17 May&
600 & 3.55 \cr
& $\pm$0.38 & $\pm$0.02 & $\pm$0.01 & $\pm$0.01 & $\pm$0.03 \cr
SDSSp J143055.90+001352.2   & 24.14 & 24.94 & 22.95 & 20.43 & 18.61 & 19 May&
1465 & 0.00 \cr
& $\pm$0.36 & $\pm$0.62 & $\pm$0.26 & $\pm$0.04 & $\pm$0.04 \cr
SDSSp J144758.46$-$005055.4 & 23.95 & 21.04 & 19.57 & 19.35 & 19.15 & 18 May&
600 & 3.78 \cr
& $\pm$0.18 & $\pm$0.03 & $\pm$0.02 & $\pm$0.02 & $\pm$0.05 \cr
SDSSp J162448.30$-$003839.4 & 22.12 & 20.36 & 20.09 & 20.15 & 20.04 & 23 Apr&
1200 & 2.92 \cr
& $\pm$0.12 & $\pm$0.02 & $\pm$0.02 & $\pm$0.02 & $\pm$0.03 \cr
\noalign{\medskip\hrule}}
\medskip
$^a$ Photometry is reported in terms of asinh magnitudes; see text
and Lupton, Gunn, and Szalay~(1999a) for details.
$^b$ Coordinate equinox is J2000.
}$$
\vskip .25truein
$$\vbox{\hsize=5truein
\halign{\hskip 12pt
# \hfil \tabskip=1em plus1em minus1em&
\hfil # \hfil &
\hfil # \hfil &
\hfil # \hfil &
\hfil # \hfil \cr
\multispan5{\hfil TABLE 2. Properties of SDSS Quasars \hfil}\cr
\noalign{\bigskip\hrule\smallskip\hrule\medskip}
\hfil Object \hfil&\hfil $z$ \hfil & \hfil $E(B-V)$ \hfil 
&\hfil $AB_{1450}$ \hfil & \hfil $M_B^{\ a}$ \hfil \cr
\noalign{\medskip\hrule\bigskip}
SDSSp J131052.52$-$005533.4 & 4.15 $\pm$ 0.01 & 0.025 & 18.98 & $-27.4$ \cr
SDSSp J134755.68+003935.1   & 3.80 $\pm$ 0.04 & 0.030 & 19.56 & $-26.7$ \cr
SDSSp J140554.07$-$000037.0 & 3.55 $\pm$ 0.03 & 0.044 & 18.70 & $-27.4$ \cr
SDSSp J144758.46$-$005055.4 & 3.78 $\pm$ 0.03 & 0.046 & 19.50 & $-26.8$ \cr
SDSSp J162448.30$-$003839.4 & 2.92 $\pm$ 0.01 & 0.092 & 20.28 & $-25.6$ \cr
\noalign{\medskip\hrule}}
\medskip
$^a$ Calculated assuming $H_0$ = 50, $q_0$ = 0.5, and $\alpha$ = $-0.5$.
}$$
\end{figure}

\acknowledgments

The Sloan Digital Sky Survey (SDSS) is a joint project of the University of
Chicago, Fermilab, the Institute for Advanced Study, the Japan Participation
Group, the Johns Hopkins University, the Max-Planck-Institute for Astronomy,
Princeton University, the United States Naval Observatory, and the University
of Washington.  Apache Point Observatory, site of the SDSS, is operated by
the Astrophysical Research Consortium.  Funding for the project has been
provided by the Alfred P.~Sloan Foundation, the SDSS member institutions,
the National Aeronautics and Space Administration, the National Science
Foundation, the U.S.~Department of Energy, and the Ministry of Education
of Japan.  The SDSS Web site \hbox{is {\tt http://www.sdss.org/}.}

The Hobby-Eberly Telescope (HET) is a joint project of the University of Texas
at Austin,
the Pennsylvania State University,  Stanford University,
Ludwig-Maximillians-Universit\"at M\"unchen, and Georg-August-Universit\"at
G\"ottingen.  The HET is named in honor of its principal benefactors,
William P. Hobby and Robert E. Eberly.  
The Marcario LRS was constructed by the University of Texas at Austin,
Stanford University, Ludwig Maximillians-Universit\"at M\"unchen, the
Instituto de Astronomia de la Universidad Nacional Autonomia de Mexico,
Georg-August-Universitaet Goettingen, and Pennsylvania State University. The
LRS is named for Mike Marcario of High Lonesome Optics who fabricated 
several optics for the instrument but died before its completion.
This work was supported in part by National Science Foundation grants
AST95-09919 and AST99-00703~(DPS), and AST96-18503~(MAS).
MAS and XF acknowledge
additional support from Research Corporation, the Princeton University
Research Board, and an Advisory Council Scholarship.

%
%
\end{document}